\documentclass[prb,twocolumn,aps,superscriptaddress,showpacs]{revtex4}

\usepackage{graphicx}
\usepackage{dcolumn}
\usepackage{amssymb}
\usepackage{bm}
\usepackage{epsfig}
\usepackage{color}

\newcommand{\CC}{{\cal C}}
\newcommand{\CGF}{\chi}

\begin{document}

\title{Nonlinear effects of phonon fluctuations on transport through nanoscale junctions}

\author{D.~F.~Urban}
\email{urban@physik.uni-freiburg.de}
\affiliation{Physikalisches Institut, Albert-Ludwigs-Universit\"at, 79104 Freiburg, Germany}
\affiliation{Departamento de F\'isica de la Materia Condensada C-XII, Facultad de Ciencias,
Universidad Aut\'onoma de Madrid, E-28049, Madrid, Spain}

\author{R. Avriller}
\affiliation{Departamento de F\'isica de la Materia Condensada C-XII, Facultad de Ciencias,
Universidad Aut\'onoma de Madrid, E-28049, Madrid, Spain}

\author{A. Levy Yeyati}
\affiliation{Departamento de F\'isica de la Materia Condensada C-XII, Facultad de Ciencias,
Universidad Aut\'onoma de Madrid, E-28049, Madrid, Spain}

\date{\today}


\begin{abstract}
We analyze the effect of electron-phonon coupling on the
full counting statistics of a molecular junction beyond
the lowest order perturbation theory. Our approach allows
to take into account analytically the feedback between
the non-equilibrium phonon and electronic distributions
in the quantum regime. We show that for junctions
with high transmission and relatively weak electron-phonon
coupling this feedback gives rise to increasingly higher
nonlinearities in the voltage dependence of the cumulants
of the transmitted charges distribution.
\end{abstract}

\maketitle

Single molecule junctions and atomic chains suspended between metallic
electrodes constitute a fascinating playground to explore the
interplay between electronic and vibronic degrees of freedom, see e.g.\ Refs.\ [\onlinecite{E-ph_Experiments_Molecules,E-ph_Experiments_AtomicWires,AbinitioCalc_el-ph_int,Viljas2005,delaVega2006}].
The interest is now not only restricted to the understanding of the mean current-voltage
characteristics but has been extended to noise properties \cite{Theory_current_shot-noise}
and, more generally, to the full counting
statistics (FCS) of the transmitted charges.\cite{E-ph_FCS_Andrei_Remi,E-ph_FCS_Federica}
An intense theoretical activity has been focussed on the analysis of the simplest model consisting
in a resonant level coupled to a single phonon mode in the quantum coherent regime.\cite{E-ph_FCS_Andrei_Remi,E-ph_FCS_Federica,Egger_Gogolin,Model_Hamiltonian_el-ph_int,Non-equ_electr_structure_vertex_corrections}
So far, however, several aspects of this problem remain to be clarified.
A serious limitation of existing transport theories is that they do not
take into account the influence of the non-equilibrium phonon
fluctuations in the statistics of the transmitted electrons, namely the
feedback of the phonon dynamics on the current-noise properties.\cite{footnoteRPA}
This limitation is associated with the breakdown of perturbation theory beyond the lowest order in electron-phonon  (e-ph) coupling as reported in Refs.\ [\onlinecite{Breakdown_Migdal_Theory}]. These works, which demonstrate the necessity of including non-perturbative effects in the analysis,  are, however, limited to equilibrium properties of bulk materials or to individual molecules in the sequential tunneling regime.

In this work we demonstrate, by a partial resummation of the perturbative expansion, the great impact of the feedback of the phonon dynamics on the quantum transport properties through nanoscale junctions with
high transmission and relatively weak e-ph coupling.
The actual signatures of phonon fluctuations
result from the interplay of several energy scales, \textit{i.e.} the
tunneling rate $\Gamma$ of electrons, the e-ph coupling $\lambda$, the phonon
frequency $\omega_0$ and the relaxation rate $\propto\eta$ of the local phonon mode due to the
coupling with bulk phonons. Depending on the specific sample considered
and the efficiency of the relaxation mechanism for the phonon population, one might obtain a regime
characterized by a thermal phonon population when $\eta \gg \lambda^2\omega_0/\Gamma^2$
(equilibrated phonons) or a regime where a strong non-equilibrium population is
generated when $\eta \ll \lambda^2\omega_0/\Gamma^2$ (unequilibrated phonons).
We analyze the crossover between the two regimes
and demonstrate that in the regime of unequilibrated phonons the electronic
current-noise shows a strong nonlinear behavior as a function of the applied
voltage for $V>\omega_0$. We attribute these nonlinearities
to a dynamical feedback of electronic quantum fluctuations that strongly renormalize the
parameters describing the local phonon mode.
This mechanism affects also the higher-order cumulants $C_n$ of the current operator which exhibit
growing nonlinearities with the order of the considered cumulant.
In particular we find the scaling
\begin{eqnarray}
    \delta C_{n+1}/\delta C_n\sim V/\omega_0
\end{eqnarray}
for the $\lambda$-dependent part $\delta C_n$ of the cumulants which is valid for $\Gamma\gg V\gg\omega_0\gg\eta$.
Despite the experimentally weak e-ph coupling
these strong nonlinearities become sizable in the regime
corresponding to recent experiments on transport through small molecules
using mechanically controllable break
junctions.\cite{E-ph_Experiments_Molecules}
Therefore they are expected to be essential to correctly capture the noise
properties above the inelastic threshold.

\begin{figure}
   \begin{center}
        \includegraphics[width=\columnwidth]{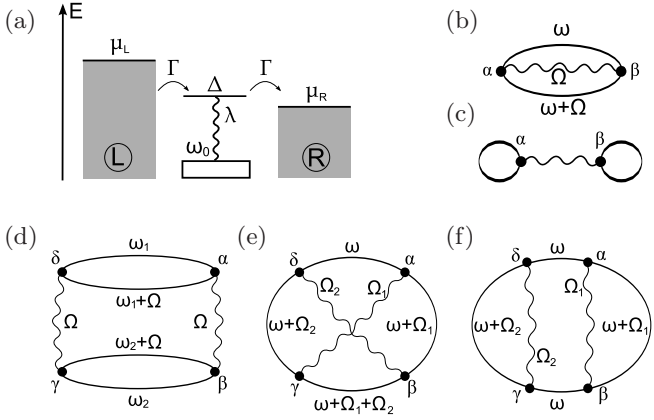}
        \caption{(a) Sketch of the model system. (b) Exchange(Fock) and (c) Hartree diagram,
        both appearing in the linked cluster expansion of the CGF at order $\lambda^2$.
        Bottom: fourth order diagrams, \emph{i.e.} (d) double-bubble, (e) double-exchange, and (f) rainbow diagram.
        \label{fig:diagrams}}
   \end{center}
\end{figure}

For our study, we use the model sketched in Fig.\ \ref{fig:diagrams}(a)
corresponding to a Hamiltonian
\begin{equation}
    H=\Delta d^\dagger d+\omega_0a^\dagger a+H_L+H_R+V_T+V_{e-ph}.
\end{equation}
Here $d$ is the annihilation
operator for the electron on the dot level at energy $\Delta$ and $a$ is the annihilation
operator for a phonon of energy $\omega_0$ (we use $e=\hbar=1$).
The left and right electrodes, described by $H_{L/R}$,
are modeled as noninteracting fermionic continua with respective field operators $\Psi_{L/R}(x)$ and the
corresponding electronic density of states $\rho_0$ is assumed to be energy independent.
The electrodes are held at different chemical potentials giving rise to
a voltage drop $V>0$. Electrons are allowed to tunnel between the
electrodes and the dot, described by the local tunnel
Hamiltonian $V_T=d^\dagger (\gamma_L \Psi_L(0) +
\gamma_R\Psi_R(0))+\mbox{h.c.}$ with left(right) hopping amplitude $\gamma_{L(R)}$.
Finally, the e-ph interaction is given in terms of the displacement operator $\Phi=a+a^\dagger$ by
$V_{e-ph}=\lambda \Phi d^\dagger d$.
In the present work we concentrate on the regime $\Gamma_{L,R}=\pi\rho_0\gamma_{L,R}^2 \gg V, \omega_0, \Delta, \lambda$ which is relevant for experiments involving small molecules\cite{E-ph_Experiments_Molecules} and chains of atoms.\cite{E-ph_Experiments_AtomicWires}
The analysis will mainly address the case of perfect transmission ($\Gamma_L=\Gamma_R=\Gamma/2$ and
$\Delta \sim 0$) where the feedback mechanism can be explained in terms of simple analytical expressions.

The FCS of charge transport through the dot is derived by calculating the cumulant generating function (CGF) $\CGF(\xi)=\CGF_0(\xi)+\delta\CGF(\xi)$
of the corresponding probability distribution $P(Q)$ to transfer the charge $Q$
during the measuring time $t_0$. The CGF of the
non-interacting system $\CGF_0(\xi)$ is given by the
Levitov-Lesovik result.\cite{Electron_Counting_Statistics,FCS_Anderson_impurity_model}
The correction due to finite e-ph coupling is given by
\begin{equation}
\label{eq:CGF}
    \delta\CGF(\xi)=-\ln\left\langle T_\CC\exp\left(-i\int_\CC dt V_{e-ph}(t)dt\right)\right\rangle_0.
\end{equation}
Here the expectation value
is taken with respect to the non-interacting Hamiltonian $H_0=H-V_{e-ph}$
and integration runs over the Keldysh contour $\CC$ with time ordering $T_\CC$
along $\CC$. The counting field $\xi(t)$ is included in the tunnel Hamiltonian $V_T$ through the
substitution $\tilde{\gamma}_L=\gamma_L e^{-i\xi(t)/2}$,
taking the value $+(-)\xi$ on the forward(backward) branch of $\CC$.
The CGF allows to access the cumulants $C_n$
of $P(Q)$ by successive derivation with respect to
the counting field,
$ C_n = \frac{(-i)^n}{t_0}\frac{\partial^n \chi}{\partial \xi^n}|_{\xi=0}$.
%

The linked cluster expansion of $\delta\chi$ in the e-ph coupling $\lambda$
can be evaluated in terms of the noninteracting electron Green function (GF)
$G_0(t,t') \,=\,-i \big{\langle} T_\CC d(t) d^{\dagger}(t')
\big{\rangle}_{0}$ and phonon propagator $D_0(t,t') \,=\,-i \big{\langle} T_\CC \Phi(t)
\Phi(t') \big{\rangle}_{0}$. The Keldysh components ($\alpha,\beta=\pm$) of the latter read
\begin{eqnarray}
    D_0^{\alpha\beta}(\omega)&=&\frac{1}{\omega-\alpha\omega_0+i\eta}-\frac{1}{\omega+\beta\omega_0-i\eta},
\end{eqnarray}
which includes the finite phenomenological parameter $\eta$ describing the coupling with bulk phonons.
The first term contributing to the expansion of $\delta\chi$ is $\propto\lambda^2$ and involves the
calculation of a Hartree-type and an exchange(Fock)-type diagram, c.f.\ Fig.\ \ref{fig:diagrams}(b)-(c).
The latter term was recently discussed in detail \cite{E-ph_FCS_Andrei_Remi,E-ph_FCS_Federica}
and was shown to be responsible for the jump in the derivative with respect
to voltage of the cumulants near the inelastic threshold at $V=\omega_0$.

On the other hand, the Hartree diagram leads to a shift of the dot level
position $\Delta$ and this renormalization is almost independent of voltage.
This statement is found to hold also for the Hartree-type diagrams of higher order
in $\lambda$ within the parameter range discussed in this article.

The next order term $\lambda^4\delta\CGF_4(\xi)$
involves the evaluation of three diagrams, which are shown in the
lower panel of Fig.~\ref{fig:diagrams}. The nomenclature \emph{double bubble} (DB),
\emph{double exchange} (DX) and \emph{rainbow} (RB) refers to the Feynman graphs of
the corresponding self-energies. In the limit of $\Gamma\gg V,\omega_0$ and at zero
temperature, we can give analytical expressions for the DX and RB diagrams and find
that they are of order $\lambda^4/\Gamma^4$. They can thus be safely neglected in
comparison to the dominant DB diagram $\delta\CGF_4^{(DB)}(\xi)\propto \lambda^4/(\Gamma^2\eta^2)$
which diverges in the limit of vanishing external damping $\eta$.

With the corresponding diagrams of higher order in $\lambda$ all being found to be divergent, we suggest
a resummation of this subset of diagrams. The Dyson equation $\hat{D}=\hat{D}_0+\hat{D}_0\hat{\Sigma}_{ph}\hat{D}$
for this RPA-type dressing of the phonon GF involves the $\xi$- and
$V$-dependent phonon self-energy
\begin{equation}
    \hat{\Sigma}_{ph}(\omega)=-i\lambda^2
    \left(\begin{array}{cc}
        \Pi^{++}(\omega)& -\Pi^{+-}(\omega) \\
        -\Pi^{-+}(\omega)& \Pi^{--}(\omega)
    \end{array}\right),
\end{equation}
where $\hat{\Pi}(\omega)$ is the charge polarization loop,
\begin{eqnarray}
     \Pi^{\alpha\beta}(\omega)&=&\int\frac{d\varepsilon}{2\pi}G^{\alpha\beta}_0(\omega+\varepsilon)G^{\beta\alpha}_0(\varepsilon).
\end{eqnarray}
Note that in the parameter range
discussed in the following, \emph{i.e.} $\lambda^2/\omega_0\Gamma \le 0.25$, we found
the RPA approximation to be charge conserving up to the chosen numerical accuracy.
The main result of the resummation is a shift and a
broadening of the phonon mode by adding a finite lifetime due to the coupling
to the electrons. A crucial aspect is to realize that this
mode renormalization is $\xi$-dependent, \emph{i.e.} it is sensitive to
the counting-field of the tunneling electrons. This has important consequences as
discussed below.
In the limit $\Gamma \gg V,\omega$ we can expand
\begin{eqnarray}
    \Pi^{-+}(\omega)&\!\!\simeq\!\!&\frac{1}{\pi\Gamma^2}\left\{\!\omega\theta(\omega)-
    \sum_{\sigma=\pm}e^{i\sigma\xi}\frac{\sigma V\!-\!\omega}{2}\theta(\omega\!-\!\sigma V)\!\right\}\quad
\\
    \Pi^{\alpha\alpha}(\omega)&\!\!\simeq\!\!&\frac{\alpha}{i\pi\Gamma}+\frac{2|\omega|+(V-|\omega|)\theta(V-|\omega|)}{2\pi\Gamma^2}
\end{eqnarray}
and thereby find four poles $\Omega_i$ of $D^{\alpha\beta}(\omega)$.
For $V>\omega_0$, the two poles
having a positive real part read
\begin{eqnarray}
\label{eq:poles}
    \Omega_{1/2} &\!=\!& (\omega_0\pm i\eta)- \frac{\lambda^2}{\pi\Gamma}- \frac{\lambda^4}{2\pi^2\Gamma^2\omega_0}
\\
    &\!\!\!\!\!\!\!\!\!\!\!\!\!\!\!\pm&\!\!\!\!\!\!\!\!
    \frac{i\lambda^2}{2\pi\Gamma^2}
    \!\left[e^{-i\xi}(V\!-\!\omega_0\!\mp\!i\eta)-(V\!+\!\omega_0\!\pm\! i\eta)\right]\!+
    {\cal O}(\Gamma^{-3}),
\nonumber
\end{eqnarray}
and due to symmetry $\Omega_{3/4}=-\Omega_{1/2}$.

\begin{figure}
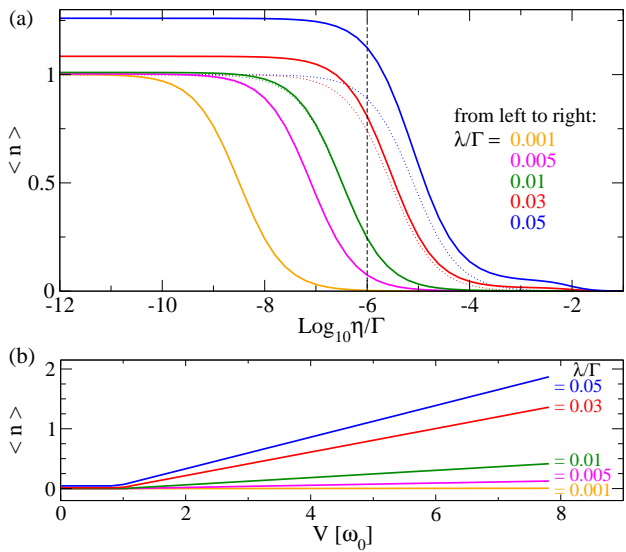

  \begin{center}
       \includegraphics[width=0.95\columnwidth]{figure2a.eps}\\
       \includegraphics[width=0.95\columnwidth]{figure2b.eps}
       \caption{(a) Mean number of phonons $\langle n\rangle$ at $V=5\omega_0$ and $\omega_0/\Gamma=0.01$
       as a function of $\eta$ shown for various values of $\lambda$.
       For comparison, the dotted curves show the results of the rate-equation approach, Eq.\ (\ref{eq:RE}), for the same parameters.
       (b) Voltage dependence of $\langle n\rangle$ for $\eta/\Gamma=10^{-6}$ and $\omega_0/\Gamma=0.01$.
       \label{fig:phononnumber}}
  \end{center}
\end{figure}

This phonon renormalization goes beyond the widely used approximation of the
non-equilibrium phonon population based on a rate equation (RE) description.\cite{AbinitioCalc_el-ph_int,Viljas2005,E-ph_FCS_Federica}
Within this RE approximation, the mean number of phonons $\langle n\rangle$ at zero temperature is given by\cite{Viljas2005}
\begin{eqnarray}
\label{eq:RE}
    \langle n\rangle_{RE}= \frac{\lambda^2 \omega_0}{4\pi\eta \Gamma^2 +
    4\lambda^2\omega_0} \left(\frac{|V|}{\omega_0} - 1\right)
    \theta(|V|-\omega_0),
\end{eqnarray}
which qualitatively describes the crossover between the equilibrated and
unequilibrated phonon regimes (for $\eta \gg \lambda^2 \omega_0 /\Gamma^2$
and $\eta \ll \lambda^2 \omega_0 /\Gamma^2$, respectively).
In Fig.\ \ref{fig:phononnumber}(a) we compare this prediction (dotted lines) with the results
for $\langle n\rangle$ obtained within the renormalized perturbation theory (solid lines)
for $\omega_0 = 0.01 \Gamma$ and different $\lambda$ values. Therefore we compute $\langle n\rangle$ by
integrating the non-diagonal component of the dressed phonon propagator,
$\langle n\rangle = - \frac{1}{2}+\frac{i}{4\pi}\int d\omega\, D^{+-}(\omega)|_{\xi=0}$.
%
%
While there is a good agreement of the RE approach with our calculations
when $\lambda < \omega_0$, discrepancies arise when
$\lambda > \omega_0$, especially in the regime of unequilibrated
phonons. The RE approach then underestimates the phonon population at a given
voltage due to the neglect of the mode renormalization effects.

\begin{figure}
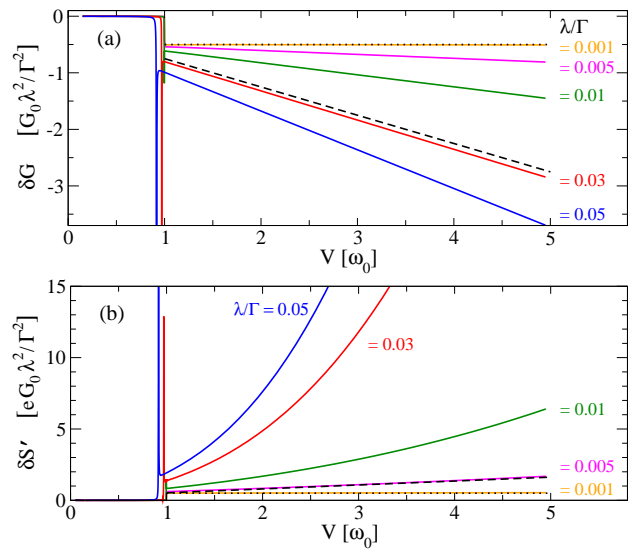

  \begin{center}
       \includegraphics[width=0.95\columnwidth]{figure3a.eps}
       \includegraphics[width=0.95\columnwidth]{figure3b.eps}
       \caption{(a) Differential conductance and (b) derivative of the current
       noise with respect to voltage as function of voltage for various values of $\lambda$.
       Parameters used are $\omega_0/\Gamma=0.01$ and $\eta/\Gamma=10^{-6}$. The dotted (dashed) lines show the results
       of the rate equation approach for thermally equilibrated (unequilibrated) phonons for comparison, taken from
       Ref.\ [\onlinecite{E-ph_FCS_Federica}].
       }
       \label{fig:transport}
  \end{center}
\end{figure}

We can now turn to the calculation of transport properties. We evaluate $\delta \CGF(\xi)$
using the dressed phonon GF,
\begin{equation}
\label{eq:deltaS}
    \delta \CGF(\xi)\simeq \frac{t_0 \lambda^2}{2i}\!\sum_{\alpha,\beta=\pm}(\alpha\beta)\int\frac{d\omega}{2\pi}\Pi^{\alpha\beta}(\omega;\xi)D^{\beta\alpha}(\omega;\xi),
\end{equation}
and from this deduce the current-voltage characteristics $I(V)$ as well as the current
noise power (second cumulant $C_2$).
Figure \ref{fig:transport} shows representative plots for the $\lambda$-dependent
contributions $\delta G$ and $\delta S'$ to the differential conductance $G=dI/dV$ and derivative of
the current noise with respect to voltage $S'=dC_2/dV$, respectively,  in the crossover regime
$\eta\sim\lambda^2\omega_0/\Gamma^2$. The obtained curves change dramatically with the
e-ph coupling. The conductance (Fig.\ \ref{fig:transport}a)
shows a negative jump at $V=\omega_0$ discussed in detail in recent works\cite{E-ph_FCS_Andrei_Remi,E-ph_FCS_Federica,Egger_Gogolin}
which is related to enhanced inelastic backscattering processes.
Note that the rounded spikes at the threshold are reminiscent of the
log-singularities obtained without the resummation which are pronounced only for
the largest $\lambda$ values shown.\cite{footnote} For voltages above the inelastic
threshold, $\delta G(V)$ is almost linear and exhibits a
crossover from a flat voltage-independent regime at low $\lambda/\Gamma=0.001$
(equilibrated phonons) to a negative slope at larger $\lambda/\Gamma=0.05$
(unequilibrated phonons). The corresponding limiting cases are plotted
(dotted and dashed lines, respectively) using the RE approach of Ref.\ [\onlinecite{E-ph_FCS_Federica}]
and the shape of the $\delta G(V)$ curves is well reproduced by this approximation.
The linear voltage dependence is directly related to the
linear population of the phonon mode for $V>\omega_0$, c.f.\ Fig.\ \ref{fig:phononnumber}(b),
and the negative slope is due to activated absorption processes
resulting from the increasing population of the local vibrational mode.

The derivative of the current noise with respect to voltage
(Fig.\ \ref{fig:transport}b) exhibits a positive jump at $V=\omega_0$ as a
result of the same inelastic backscattering mechanism responsible for the
jump in $\delta G(V)$. However, for $V>\omega_0$ the behavior of the curves
is qualitatively different. In the regime of equilibrated phonons, $\delta S'(V)$
is almost voltage-independent and well reproduced by the RE approach. For
unequilibrated phonons however, the current-noise curves develop strong
nonlinearities which are absent in the RE approach. We stress that already
for quite small $\lambda/\Gamma=0.005$ (corresponding to
an almost thermally equilibrated regime, c.f.\ Fig.\ \ref{fig:phononnumber}a)
the amplitude of the nonlinearities is large enough to match the (maximal) results of
the RE approach for unequilibrated phonons (dashed line in Fig.\ \ref{fig:transport}b).
This qualitative different behavior can be understood as follows.
Within the RE approach and in the limit $\Gamma \gg
\omega_0, V, \lambda$ in which we are interested, the correction to
the FCS due to e-ph coupling is purely Poissonian,\cite{E-ph_FCS_Andrei_Remi,E-ph_FCS_Federica}, i.e.
$\delta C_{n,RE} \propto \delta C_{1,RE}$. On the other hand, the
correction to the FCS in the renormalized theory departs from being
Poissonian due to the dynamical renormalization of the phonon mode.
In the particular regime of unequilibrated phonons ($\eta\ll\lambda^2\omega/\Gamma^2$)
and for $\Gamma\gg V>\omega_0$,
the dressed phonon propagator can be approximated by
\begin{equation}
\hat{D}\simeq\left(\prod_{j=1}^4\frac{\sqrt{2\omega_0}}{\omega-\Omega_j}\!
    \right)\!\!
    \left(\!\!\begin{array}{cc}
        f-i\lambda^2\Pi^{--}& -i\lambda^2\Pi^{+-} \\
        -i\lambda^2\Pi^{-+}& -f-i\lambda^2\Pi^{++}
    \end{array}\!\!\right)
\end{equation}
with $f=(\omega^2-\omega_0^2)/2\omega_0$ and the poles $\Omega_j$ given in Eq.\ (\ref{eq:poles}).
The integral of Eq.\ (\ref{eq:deltaS}) is then approximated by
\begin{eqnarray}
    \delta \CGF(\xi)&\simeq&\frac{t_0 \lambda^2}{2i}
    \sum_{j=1}^4
    \frac{4\omega_0^2}{\prod_{k\neq j}(\Omega_j-\Omega_k)}
    \sum_{\alpha,\beta=\pm}\!(\alpha\beta)
 \nonumber    \\
  &&\!\!\!\!\!
    \int\!\!\frac{d\omega}{2\pi}\frac{\Pi^{\alpha\beta}(\omega)
    \left[\alpha f(\omega)\delta_{\alpha\beta}-i\lambda^2\Pi^{-\alpha,-\beta}(\omega)\right]}{\omega-\Omega_j},\nonumber\\
\end{eqnarray}
where the prefactor can be expressed as
\begin{equation}
  \frac{4\omega_0^2}{\prod_{k\neq j}(\Omega_j-\Omega_k)}\simeq\sigma_{\!j}\frac{i\pi\Gamma^2}{\lambda^2\omega_0}\frac{e^{i\xi}}{1+e^{i\xi}}
  \sum_{n=0}^{\infty}\!\left[-i\tan\!\left({\xi}/{2}\right)\frac{V}{\omega_0}\right]^n
\end{equation}
with $\sigma_j=-{\rm sign}({\rm Im}(\Omega_j))$. This factor therefore is identified as the source of the increasing nonlinear
voltage dependence when taking successive derivatives of $\delta\chi$ with respect to $\xi$ at $\xi=0$.
Finally, the most nonlinear contribution to the $\delta C_n$ in this regime
reads
\begin{equation}
    \delta C_n \sim (-1)^n(n!)
    \frac{2\lambda^2\omega_0}{\pi\Gamma^2}\left(\frac{V}{2\omega_0}\right)^{n+1},
\end{equation}
which demonstrates the scaling law anticipated in the introduction.\cite{footnoteVoltage}

In conclusion, we have analyzed the effect of e-ph interaction on the FCS of a
molecular junction beyond the lowest order perturbation theory. A RPA-type resummation
of diverging diagrams results in a renormalization of the phonon parameters which
depends on the electron counting-field, thereby giving rise to an increasingly higher nonlinear
voltage dependence in the cumulants of the transmitted charges distribution.
These strong nonlinear effects are expected to be
accessible in low-temperature transport experiments performed on highly conductive molecular
junctions\cite{E-ph_Experiments_Molecules} and
atomic chains,\cite{E-ph_Experiments_AtomicWires} where
typically the energy of the strongly coupled vibrational mode and the e-ph coupling
strength are of order $\omega_0/\Gamma \approx 0.01$ and $\lambda/\Gamma \approx 0.1-0.01$, respectively.

The authors would like to thank F.\ Haupt, A.\ Komnik, R.\ Egger, H.\ Grabert,
Jan van Ruitenbeek, J.C.\ Cuevas and A.\ Martin-Rodero for many interesting
and fruitful discussions.
Financial support from the Spanish MICINN under contract
NAN2007-29366-E (CHENANOM) is acknowledged.
D.\ Urban and R.\ Avriller contributed equally to this work.


\begin{thebibliography}{10}
\bibitem{footnote} Such spikes are smeared further at finite temperatures but the overall behavior of
   $\delta G$ and $\delta S'$ would not be affected provided that $T \ll \omega_0$.
\bibitem{footnoteVoltage}
Note that throughout the paper we assume $V>0$. However, $C_n(-V)=(-1)^nC_n(V)$.
\end{thebibliography}
\end{document}